\documentclass[acmtog,nonacm]{acmart}
\usepackage{pifont}
\usepackage{float}
\usepackage{gensymb}

\AtBeginDocument{%
  }

\setcopyright{acmlicensed}
\copyrightyear{2026}
\acmYear{2026}
\acmDOI{XXXXXXX.XXXXXXX}

\acmJournal{TOG}
\acmVolume{37}
\acmNumber{4}
\acmArticle{111}
\acmMonth{8}
\usepackage{wrapfig}
\usepackage{nicefrac}
\usepackage{mathtools}
\usepackage{graphicx}
\usepackage{booktabs} 
\usepackage{combelow} 
\usepackage[ruled,vlined,linesnumbered]{algorithm2e}
\usepackage{tabularx} 
\usepackage{colortbl} 
\usepackage{cancel}
\usepackage{makecell} 
\usepackage{enumitem}
\usepackage{manfnt} 
\usepackage[nameinlink]{cleveref}

\usepackage{ccicons}
\usepackage{amsmath}
\usepackage{amsfonts}
\usepackage{amsbsy}

\usepackage{listings} 
\usepackage{courier}
\definecolor{mygreen}{rgb}{0,0.6,0}
\definecolor{mygray}{rgb}{0.5,0.5,0.5}
\definecolor{mymauve}{rgb}{0.58,0,0.82}
\definecolor{superlightgray}{RGB}{240,240,240}
\definecolor{derekBlue}{RGB}{144,210,236}
\definecolor{caltechOrange}{RGB}{255,108,12}
\definecolor{iglGreen}{RGB}{153,203,67}
\definecolor{coralRed}{RGB}{250,114,104}
\definecolor{gray}{RGB}{200,200,200}

\newcommand{\refequ}[1] {Eq.~\ref{equ:#1}}
\newcommand{\reffig}[1] {Fig.~\ref{fig:#1}}
\newcommand{\reftab}[1] {Table~\ref{tab:#1}}
\newcommand{\refsec}[1] {Section~\ref{sec:#1}}

\newcommand{\refalg}[1] {Algorithm~\ref{alg:#1}}

\DeclareMathOperator*{\argmax}{arg\,max}

\newcommand{\M}{\mathcal{M}}
\newcommand{\G}{\mathcal{G}}
\renewcommand{\S}{\mathcal{S}}
\newcommand{\I}{\mathcal{I}}

\newcommand{\vecFont}[1]{\mathbf{#1}}

\def\vv{{\vecFont{v}}}

\newcommand{\matFont}[1]{\mathbf{#1}}

\def\mE{{\matFont{E}}}
\def\mF{{\matFont{F}}}

\def\mT{{\matFont{T}}}

\def\mV{{\matFont{V}}}
\def\mW{{\matFont{W}}}

\begin{document}
\title{A Geodesic Cut-Cell Prior for Neural Skinning}

\author{Wenchao Ma}
\authornote{Work partially completed during an internship at Roblox.}
\email{wmm5390@psu.com}
\affiliation{%
  \institution{Penn State University}
  \country{USA}
}
\author{Surya Dwarakanath}
\email{sdwarakanath@roblox.com}
\affiliation{%
  \institution{Roblox}
  \country{USA}
}
\author{Yizhak Ben-Shabat}
\email{sitzikbs@gmail.com}
\affiliation{%
  \institution{Roblox}
  \country{USA}
}
\author{Dario Kneubühler}
\email{dkneubuhler@roblox.com}
\affiliation{%
  \institution{Roblox}
  \country{USA}
}
\author{Haomiao Jiang}
\email{haomiaojiang@roblox.com}
\affiliation{%
  \institution{Roblox}
  \country{USA}
}
\author{Sharon X. Huang}
\email{suh972@psu.edu}
\affiliation{%
  \institution{Penn State University}
  \country{USA}
}
\author{Hsueh-Ti Derek Liu}
\email{hsuehtil@gmail.com}
\affiliation{%
  \institution{Roblox}
  \country{Canada}
}

\begin{abstract}
We introduce \emph{cut-cell skinning}, a geometric prior designed to augment data-driven skinning weight generation. 
While data-driven methods show promise in producing high-quality skinning weights, they often lack the generalizability of classic geometric approaches. To bridge this gap, we propose a geometric prior that can be robustly computed for in-the-wild meshes and is efficient for large-scale machine learning workflows. 
The key idea of our cut-cell skinning is a fast graph-based approximation of the volumetric geodesics distances, motivated by their importance in classic skinning weight computation.
Our method achieves orders of magnitude speedup compared to optimization-based solvers and remains resilient to topological artifacts common in cage- or voxel-based alternatives.
We demonstrate the efficacy of the cut-cell skinning prior by integrating it into recent neural skinning models, showing consistent improvements across existing methods and achieving state-of-the-art results. Project page: \textcolor{blue}{\href{{https://wenchao-m.github.io/CutCell.github.io/}}{https://wenchao-m.github.io/CutCell.github.io/}}

\end{abstract}

\begin{teaserfigure}
    \centering
    \includegraphics[width=\linewidth]{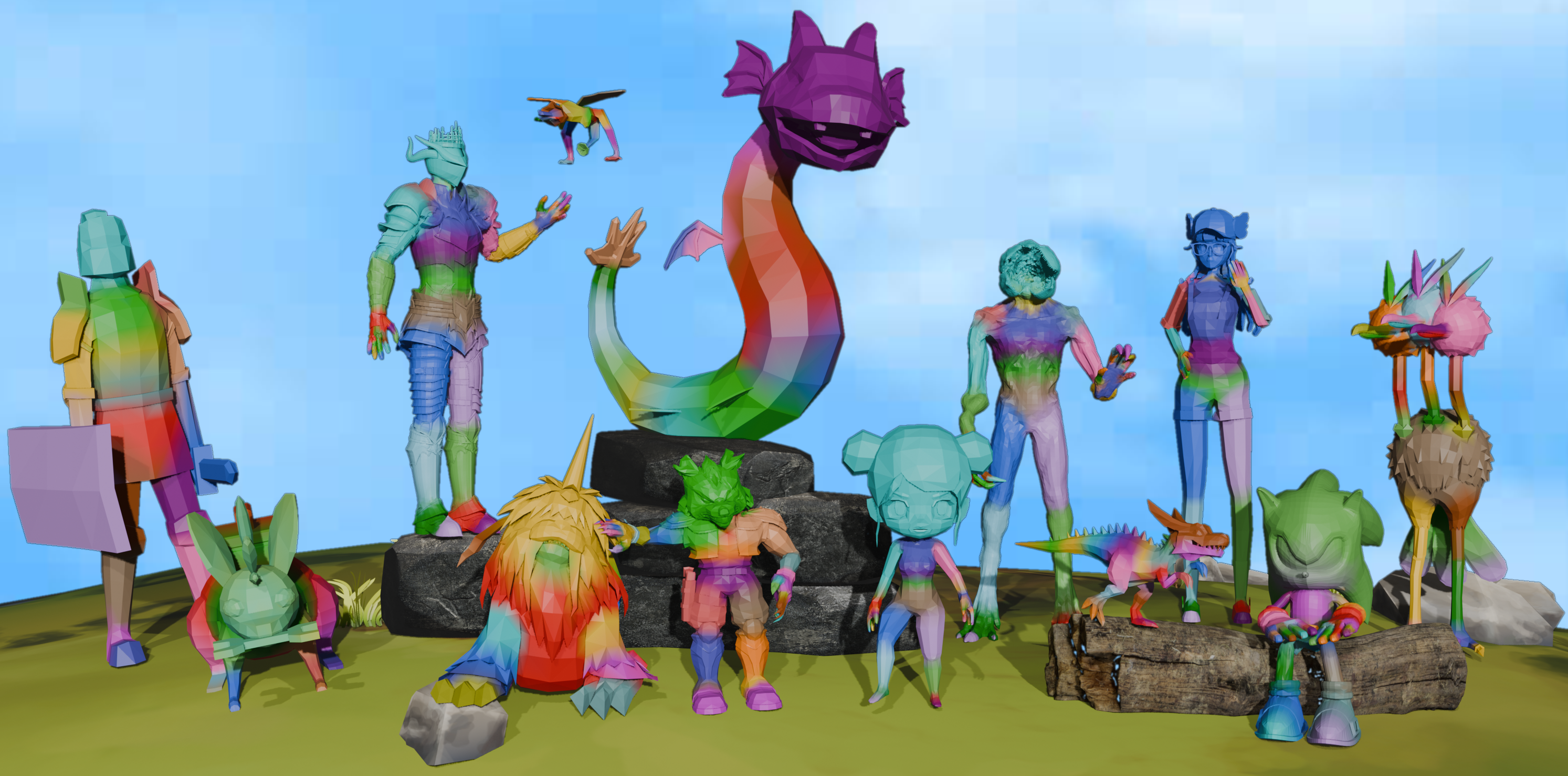}
    \caption{We show a diverse set of skinning weight predictions produced by a neural skinning method augmented with our \emph{geodesic cut-cell prior}. Our prior provides geometric guidance to complement learning-based solutions, improving their generalization across diverse geometries while maintaining their semantic understanding.}
    \label{fig:teaser}
\end{teaserfigure}

\maketitle
\section{Introduction}
\begin{figure}
    \centering
    \includegraphics[width=3.33in]{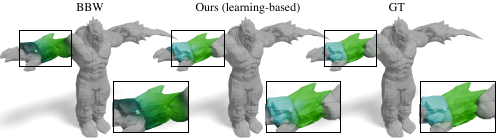}
    \caption{Artist-created skinning weights respect the semantics of the underlying geometry, such as the rocky lower arm (right). Pure geometric methods (left), such as the Bounded Biharmonic Weight (BBW)~\cite{JacobsonBPS11}, has no semantic awareness, leading to overly smoothed skinning weights without capturing material heterogeneity. Our approach (middle) combines data-driven semantic learning with a robust geometric prior, yielding skinning weights that better match the ground truth.}
    \label{fig:semantic_weights}
\end{figure} 

Linear Blend Skinning (LBS) remains the predominant paradigm for real-time character animation. At its core, LBS formulates deformation as a weighted linear combination of skeletal transformations; consequently, achieving high-fidelity deformation necessitates meticulously authored skeletal rigs and skinning weights.

Traditionally, these rigs and weights are manually ``painted'', resulting in labor-intensive processes that require domain expertise. This has catalyzed the development of automated methods that, for example, compute skinning weights by solving constrained optimization problems~\cite{JacobsonBPS11,BaranP07, WangSolomon21}. While such methods exhibit strong generalizability and satisfy fundamental skinning weight properties, such as smoothness and locality, they often fail to capture the semantic nuances required to articulate characters with material heterogeneity (see \reffig{semantic_weights}).

These limitations have motivated the exploration of machine learning alternatives. However, while they show promise in capturing semantics, they frequently struggle to generalize to out-of-distribution data and fail to maintain the desired skinning properties. This observation leads to the simple motivation of our work: can the geometric method be effectively integrated as an inductive bias to guide data-driven skinning weight inference?

Realizing this integration presents significant challenges. To enable scalable training on large-scale datasets within average computational budgets, we require a geometric approach capable of operating robustly and efficiently on arbitrary meshes. Current geometric techniques often rely on computationally intensive volumetric meshing and optimization~\cite{JacobsonBPS11,WangSolomon21}, or error-prone caging~\cite{JoshiMDGS07, JuSW05} and voxelization steps~\cite{DionneL13}, both of which are intractable for processing large-scale datasets reliably.

In lieu of this, we propose \emph{cut-cell skinning} as an efficient framework for computing geometric skinning weight priors.
Drawing inspiration from classical methods where volumetric geodesic distances are paramount, our approach employs a fast and robust approximation of volumetric geodesics via graph-based geodesics. 
In comparison to existing methods, our approach achieves a speedup of two to four orders of magnitude over optimization-based solvers (e.g., \cite{JacobsonBPS11}). Furthermore, it remains resilient to artifacts caused by proximal mesh components, a common failure mode in cage- or voxel-based alternatives (e.g., \cite{DionneL13}).
We demonstrate the efficacy of the cut-cell skinning prior by integrating it into recent neural skinning architectures with minimal architectural modifications needed to inject our prior. When combined with network inference, our method consistently yields improvements upon existing neural skinning models and achieves state-of-the-art result on benchmarks. We further propose a deformation-space evaluation metric that captures animation-time artifacts overlooked by standard weight-space metrics, under which our cut-cell prior shows the largest improvements. In addition, we analyze redundancy in the commonly used Articulation-XL 2.0 dataset and introduce a rigorously de-duplicated evaluation split to avoid unintended train–test overlap; under this stricter protocol, our method consistently outperform existing baselines. 

\section{Related Work}
Linear Blend Skinning (LBS) remains the ubiquitous standard for real-time character deformation due to its simplicity and computational efficiency~\cite{JacobsonG14}. The quality of deformation depends on the skeleton geometry and the associated skinning weights. 
Our work focuses on the weight computation problem, specifically aiming to bridge the gap between robust geometric generalizations and semantic data-driven predictions.

\subsection{Geometric Skinning Weights}
\begin{figure}
    \centering
    \includegraphics[width=3.33in]{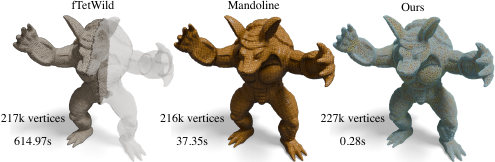}
    \caption{An alternative approach for constructing graphs for skinning weight computation is to use volumetric meshing, such as tetrahedral (fTetWild~\cite{HuSWZP20}, left) or cut-cell meshes (Mandoline~\cite{TaoBFL19}, middle). While these methods produce high-quality volumetric representations, their high construction cost makes them impractical for large-scale datasets. In contrast, our method achieves substantially lower runtime at a similar resolution.
    }
    \label{fig:meshing_time}
\end{figure}

Geometric approaches have long served as the foundation for automatic skinning. 
%
To achieve high-quality, smooth weights, methods like Bounded Biharmonic Weights (BBW)~\cite{JacobsonBPS11} formulate weight computation as a constrained energy minimization problem. BBW produces widely used, high-quality results that respect the partition of unity. However, it requires expensive volumetric meshing and solving linear systems, making it computationally heavy for large-scale training pipelines. Recent accelerations, such as the dual-formulation solver~\cite{solomon2025computing} and Quasi-Harmonic Weights~\cite{WangSolomon21}, improve runtime significantly but still retain the dependency on expensive volumetric meshing (see \reffig{meshing_time}).
They offer strong generalizability across unseen shapes but lack the semantic understanding required for complex characters (e.g., distinguishing a rigid shell from soft skin).

\begin{figure}
    \centering
    \includegraphics[width=3.33in]{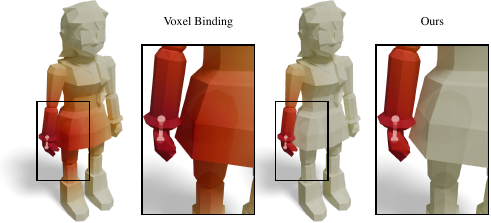}
    \caption{We compare the geodesic distances from the bone to the mesh vertices using a voxelization-based method \cite{DionneL13} and our approach. Voxelization may introduce unintended connections across narrow gaps between nearby surface regions (e.g., the hand and the torso), leading to artifacts in the geodesic distance, whereas our cut-cell graph preserves separation across these regions and produces more accurate distance estimates.}
    \label{fig:voxelBinding}
\end{figure}

To avoid the fragility of tetrahedral meshing, industry pipelines often rely on voxel-based approximations. Geodesic Voxel Binding~\cite{DionneL13} computes weights via geodesic distances over a voxelized domain. This approach is highly robust to imperfect geometry but introduces discretization artifacts and often merges spatially close but geodesically distant parts (e.g., binding legs together with arms in \reffig{voxelBinding}). 
\begin{figure}
    \centering
    \includegraphics[width=3.33in]{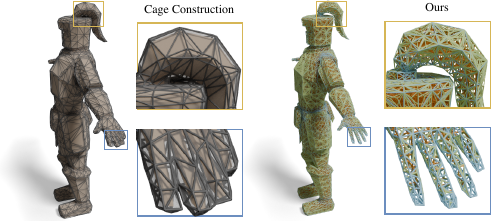}
    \caption{Cage-based methods propagate bone influence through an enclosing cage, which is prone to merge close but distinct parts (left). Our cut-cell graph preserves correct separation between nearby mesh parts (right).}
    \label{fig:cage}
\end{figure} 

Other coordinate-based methods avoid voxelization but introduce their own topological constraints: Harmonic Coordinates~\cite{JoshiMDGS07} and Mean Value Coordinates~\cite{JuSW05} require watertight bounding cages with correct cage topology (see \reffig{cage}), while Heat Diffusion~\cite{BaranP07} and Biharmonic weights (without bounds)~\cite{BotschK04} can produce negative values or unintuitive ``push-pull'' artifacts. 
Our work builds on the motivation that while these geometric methods are robust priors, they fundamentally lack the semantic insight that can only be resolved through learning~\cite{BangL18}.

\subsection{Data-Driven Skinning Weights}
Recent data-driven methods have sought to overcome geometric limitations by learning weight distributions directly from datasets of rigged characters. However, integrating geometric priors into these learning frameworks remains non-trivial.


Early approaches relied on handcrafted geometric features to guide learning. NeuroSkinning~\cite{liu2019neuroskinning} and RigNet~\cite{xu2020rignet} use Graph Neural Networks (GNNs) with pre-computed voxel-based geodesic distances to bones~\cite{DionneL13}. While effective, approximation errors in these distance features can limit skinning accuracy.
UniRig~\cite{zhang2025unirig} and MagicArticulate~\cite{song2025magic} continue this trend by explicitly conditioning on voxel-based blending weights as priors, acknowledging that geometry is a necessary guide for the learning process.

To bypass feature engineering, recent Transformer based architectures like RigAnything~\cite{liu2025riganything}, Puppeteer~\cite{song2025puppeteer}, and Anymate~\cite{deng2025anymate} employ attention mechanisms (e.g., joint-vertex cross-attention) to infer weights globally. While these methods achieve impressive results by learning ``soft'' associations, they often struggle to generalize to unseen topologies compared to purely geometric solvers. SkinningNet~\cite{albert2022skinningnet} attempts to bridge this by learning a connectivity graph, but fundamentally, these methods must ``re-learn'' geometric smoothness from scratch.
Our approach takes a distinct path: rather than discarding the geometric solver or treating it as a static feature, we incorporate geometric proxy as a prior to several network architectures, demonstrating the effectiveness of combining the knowledge from geometric approaches with the semantic precision of learning.

\section{Geodesic Cut-Cell Graph Skinning}
\label{sec:cutcell}

Our work focuses on demonstrating the effectiveness of incorporating geometric priors to neural skinning architectures. We first focus on the construction of our prior -- cut-cell graph skinning, followed by modifications to state-of-the-art architectures to inject the prior in \refsec{results}.

To operate on a large amount of surface mesh data in-the-wild, we desire a robust and efficient method to discretize the volume enclosed by the surface mesh and compute geometric prior where the network can learn to predict skinning weights based on it. 

These design requirements make volumetric meshing, such as Tetrahedral \cite{HuSWZP20} or Cut-Cell \cite{TaoBFL19} meshes, less attractive because processing each mesh may take minutes to hours (see \reffig{meshing_time}), leading to weeks of pre-processing time for entire dataset. 
We propose an efficient alternative -- \emph{Cut-Cell Graph} -- to approximate the true volumetric geodesic distances with graph geodesics at a substantially lower computational cost. Then we convert distances to a skinning prior with kernel functions, similar to \cite{DionneL13}. 
In a nutshell, our graph construction ``cuts'' through grid cells with the input surface, thus inheriting the terminology ``\emph{cut-cell}'' from fluid simulation \cite{berger2017cut}.

\subsection{Graph Construction}
\label{subsec:graph_construction}
\begin{figure}
    \centering
    \includegraphics[width=3.33in]{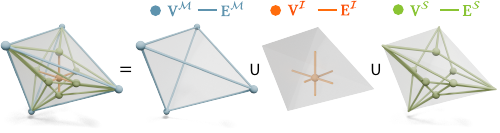}
    \caption{A cut-cell graph $\G$ consists of vertices $\mV^\M$ and edges $\mE^\M$ from the input mesh $\M$, interior vertices $\mV^\I$ and edges $\mE^\I$ from the voxel grid, and a set of surface vertices $\mV^\S$ where the mesh and the grid meets. We further augment the graph with edges $\mE^\S$ that connects $\mV^\S$ and $\mV^\M$ to ensure the graph vertices are all connected. }
    \label{fig:cutCellGraph_3D}
\end{figure}
\begin{figure}
    \centering
    \includegraphics[width=3.33in]{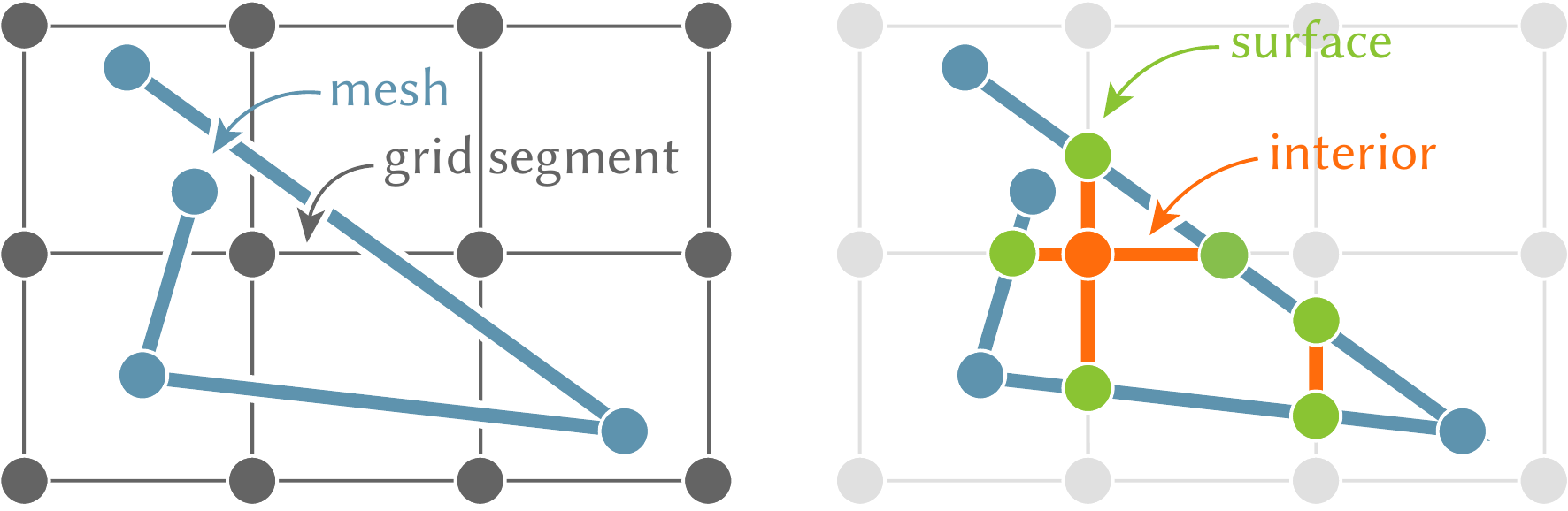}
    \caption{For each grid segment, diced by the grid vertices (black) and the mesh (blue), we use the \emph{generalized winding number} to decide whether they lie on the interior of the mesh. If they are, we keep their edges (orange) and tip vertices (green and orange) in the final cut-cell graph.}
    \label{fig:raySegments}
    \vspace{-4mm}
\end{figure}

Let $\G = (\mV, \mE)$ be the cut-cell graph with a set of graph vertices $\mV$ and edges $\mE$. 
The graph vertices $\mV = (\mV^\M, \mV^\I, \mV^\S)$ consist of the vertices from the input \emph{mesh} $\mV^\M$, a set of voxel grid vertices $\mV^\I$ on the \emph{interior} of the mesh, and a set of intersecting points between voxel grid edges and the mesh \emph{surface} $\mV^\S$.
The graph edges $\mE = (\mE^\M, \mE^\I, \mE^\S)$ consist of the mesh edges $\mE^\M$, a set of \emph{interior} voxel edges $\mE^\I$ classified as inside the mesh, and a set of \emph{surface} edges $\mE^\S$ connecting $\mV^\S$ to nearby mesh vertices $\mV^\M$.
In \reffig{cutCellGraph_3D}, we provide a simple example to illustrate these components.

%


To aim for extreme efficiency and robustness, constructing the cut-cell graph $\G$ only involves two extremely robust subroutines -- ray casting and querying the generalized winding number \cite{BarillDSLJ18} -- which can be trivially parallelized. 
The first parallelizable step is to cast axis-aligned rays to compute intersections with the input mesh $\M$. Specifically, for each axis-aligned direction $x$, we sample ray origins at 2D grid locations on the plane $x = 0$ and cast rays towards the $+x$ direction (same for $y$ and $z$), assuming the mesh is pre-normalized to the unit cube. Each ray will intersect with the triangle mesh $\M$ at multiple locations, and pass through several voxel vertices. These points will ``cut'' the ray into multiple segments, which we collect as the potential candidates for graph elements.
After collecting all segments, we perform parallel queries of the generalized winding number \cite{BarillDSLJ18} at the midpoint of each segment to classify whether this segment is inside or outside the mesh (see \reffig{raySegments}). We then keep all edge segments $\mE^\I$ on the interior of the mesh and all the vertices associated with these interior edges, including both the interior voxel vertices $\mV^\I$ and the surface vertices $\mV^\S$ .

If the mesh is watertight, we can leverage the property that the winding number field is piece-wise constant, and merging segments ``cut'' by voxel vertices. This can dramatically reduce the number of segments to query (similar to \cite{TrettnerNK22}), and further speedup the construction time.

The above process constructs a graph that fills in the interior of the mesh $M$, but does not connect to the surface mesh vertices $\mV^\M$. We leverage the fact that each intersecting vertex in $\mV^\S$ is a barycentric point on $\M$, and add extra \emph{surface} edges $\mE^\S$ connecting each intersecting point to the three corner vertices of the triangle it hits (see the green edges in \reffig{cutCellGraph_3D}). 

\begin{figure}
    \centering
    \includegraphics[width=3.33in]{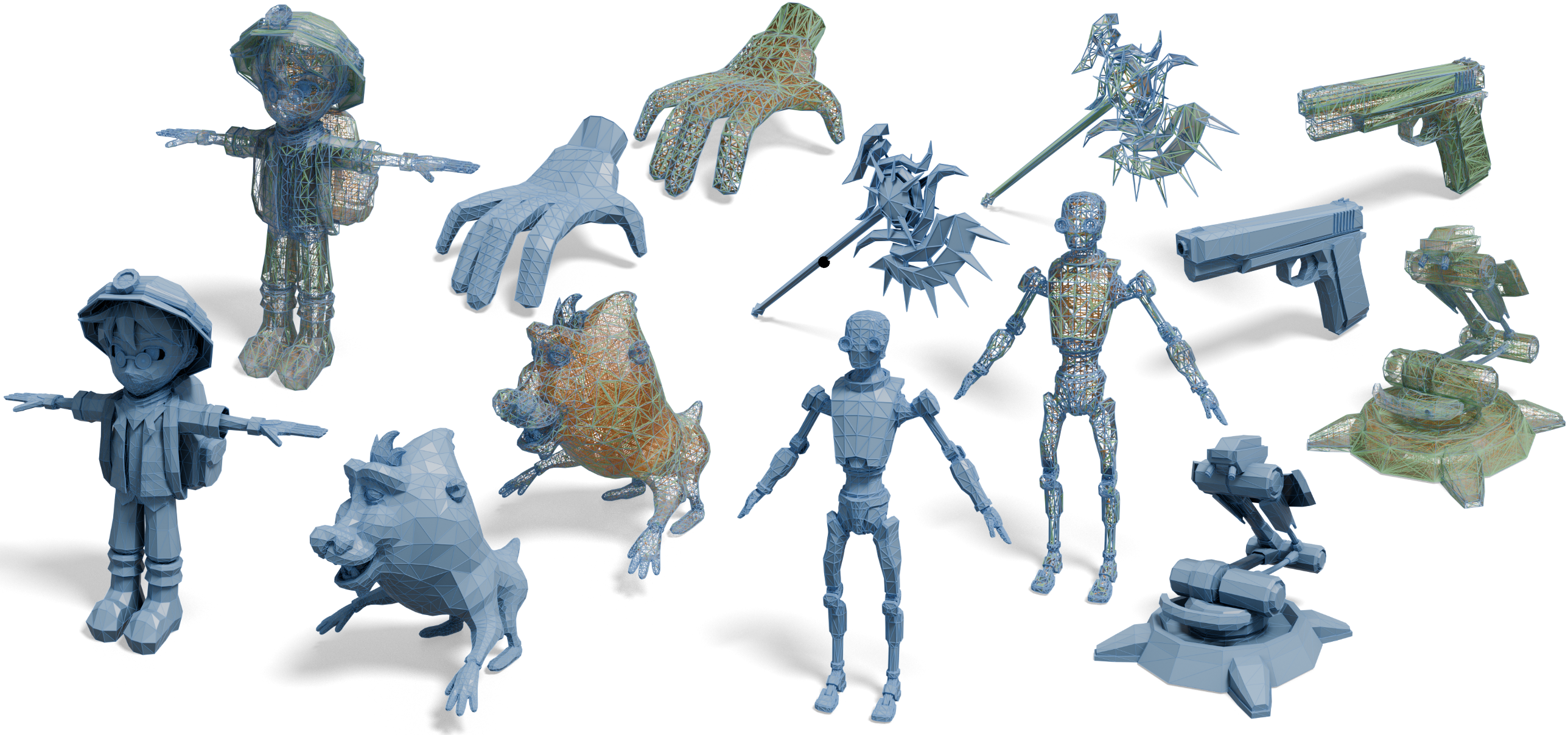}
    \caption{We show a gallery of our cut-cell graphs (right) alongside the corresponding in-the-wild input meshes (left). We use the same color scheme as Fig. 5, where the orange color denotes the \emph{interior}, the green color denotes the \emph{surface intersection}, and the blue color denotes the input \emph{mesh} vertices/edges.}
    \label{fig:cutCellGraphGallery}
    \vspace{-1mm}
\end{figure}

We additionally incorporate the original mesh vertices $\mV^\M$ and edges $\mE^\M$ to the graph $\G$ to avoid some vertices that have no connection to the graph due to insufficient grid sampling. We summarize the graph construction process in \refalg{cutcell_graph}, and show a gallery of results in \reffig{cutCellGraphGallery}.

The construction of our cut-cell graph is efficient. For a fair comparison, we demonstrate that even a baseline CPU implementation of our approach achieves orders-of-magnitude speedups over
\begin{wrapfigure}[8]{r}{1.3in}
    \vspace{-2mm}
	\includegraphics[width=\linewidth, trim={10mm 0mm 0mm 0mm}]{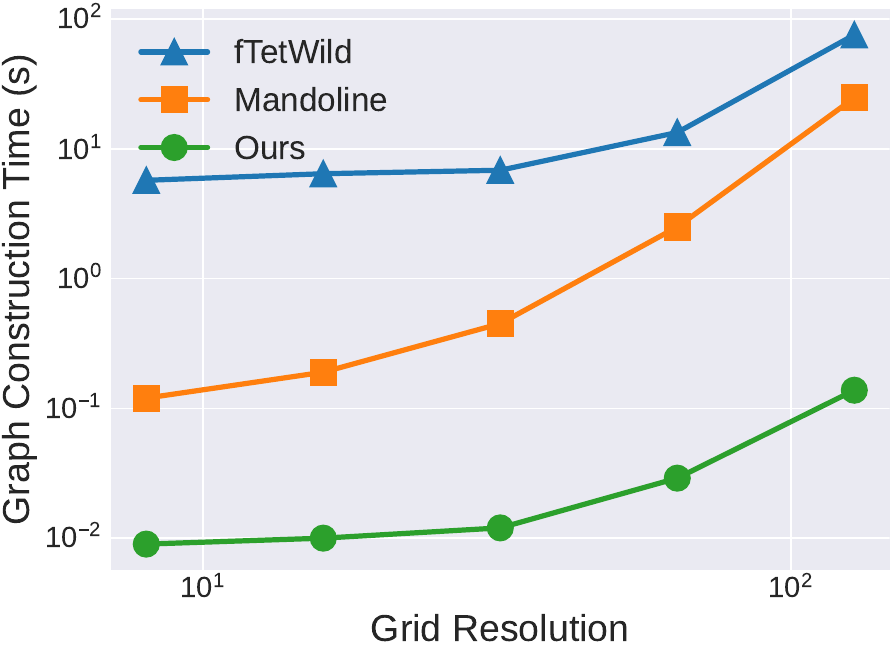}
	\label{fig:graph_time} 
\end{wrapfigure}
alternative volumetric meshing techniques (see \reffig{meshing_time} and the inset). Future work on GPU implementation to parallelize the ray tracing and winding number queries could further accelerate our method.

\begin{algorithm}[t]
\caption{Cut-Cell Graph Construction}
\label{alg:cutcell_graph}
\KwIn{Mesh $(\mV^\M,\mF^\M)$, grid resolution $R$}
\KwOut{Cut-cell graph $\mathcal{G}=(\mV,\mE)$}

Initialize $\mV^\I=\emptyset$, $\mV^\S=\emptyset$.\;
\ForEach{axis $a\in\{x,y,z\}$}{
  \tcp{Cast $R^2$ rays: one per voxel column on the plane orthogonal to $a$}
  $\mathcal{O}\leftarrow$\textsc{RayOrigins}$(R,a)$\;
  $\mathcal{I}\leftarrow$\textsc{RayMeshIntersections}$(\mV^\M,\mF^\M,\mathcal{O},a)$\;
  $\mathcal{S}\leftarrow$\textsc{BuildRaySegments}$(\mathcal{I})$\;
  $\mathcal{P}\leftarrow\{\textsc{SegmentMidpoint}(s)\mid s\in\mathcal{S}\}$\;
  $\mathcal{W}\leftarrow$\textsc{BatchWindingNumber}$(\mV^\M,\mF^\M,\mathcal{P})$\;
  \ForEach{segment $s\in\mathcal{S}$}{
    \If{$\mathcal{W}[s]\ge\tau$}{
      $\mV^\I \leftarrow \mV^\I \cup$\textsc{IncidentVoxelVertices}$(s,R)$\;
      $\mV^\S \leftarrow \mV^\S \cup$\textsc{SegmentIntersections}$(s)$\;
    }
  }
}
$\mV \leftarrow \mV^\M \cup \mV^\I \cup \mV^\S$\;
$\mE \leftarrow$\textsc{BuildVoxelEdges}$(\mV^\I)\cup$\textsc{BuildMeshEdges}$(\mF^\M)$\;
\ForEach{$x\in \mV^\S$}{
  Let $(v_1,v_2,v_3)$ be the vertices of the intersected triangle\;
  $\mE \leftarrow \mE \cup \{(x,v_1),(x,v_2),(x,v_3)\}$\;
}
Assign Euclidean edge weights $w(u,v)=\|\mathbf{p}_u-\mathbf{p}_v\|_2$\;
$\mathcal{G}\leftarrow(\mV,\mE)$;\;
\Return{$\mathcal{G}$}
\end{algorithm}

\begin{algorithm}[t]
\caption{Bone Distances on Cut-Cell Graph}
\label{alg:cutcell_bonedist}
\KwIn{Graph $\mathcal{G}=(\mV,\mE)$, bones $\mathcal{B}=\{b=(p_0,p_1)\}$, samples $N$}
\KwOut{Bone distance fields $\{D_b\}_{b\in\mathcal{B}}$}

\ForEach{bone $b=(p_0,p_1)\in\mathcal{B}$}{
  $Q\leftarrow$\textsc{SampleBonePoints}$(p_0,p_1,N)$\;
  $S_b\leftarrow$\textsc{NearestGraphVertices}$(\mathcal{G},Q)$\;
  $\delta(u)\leftarrow$\textsc{PointToSegmentDistance}$(\mathbf{p}_u,p_0,p_1)$ for $u\in S_b$\;
  $D_b\leftarrow$\textsc{MultiSourceDijkstra}$(\mathcal{G},S_b,\delta)$\;
}

\tcp{Handle unreachable vertices}
\ForEach{vertex $v\in\mV$}{
  \If{$\forall\, b\in\mathcal{B},~ D_b(v)=\infty$}{
    $\mathcal{B}_3(v)\leftarrow$\textsc{TopKNearestBones}$(v,\mathcal{B},k=3)$\;
    \ForEach{$b\in \mathcal{B}_3(v)$}{
      $D_b(v)\leftarrow$\textsc{EuclideanBoneDistance}$(v,b)$\;
    }
  }
}
\Return{$\{D_b\}$}
\end{algorithm}

\subsection{Skinning Prior}
\label{subsec:skinning_prior}
\begin{figure}
    \centering
    \includegraphics[width=3.33in]{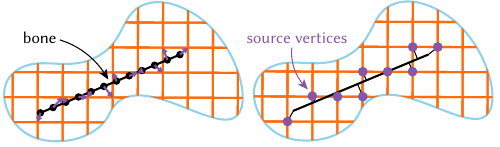}
    \caption{In order to compute graph distances to the vertices, we start by sampling $N$ points on the bone (black points, left) and compute their closest graph vertices (purple arrows, left) as a form of ``voting'' source vertices to start the distance traversal. After the voting process, we collect a set of source vertices (purple points, right) and initialize their distances as their closest Euclidean distance to the bone (black line), followed by the Dijkstra's algorithm to compute distance values to all non-source graph vertices in $\G$. }
    \label{fig:distanceInit}
\end{figure}

\begin{figure}
    \centering
    \includegraphics[width=3.33in]{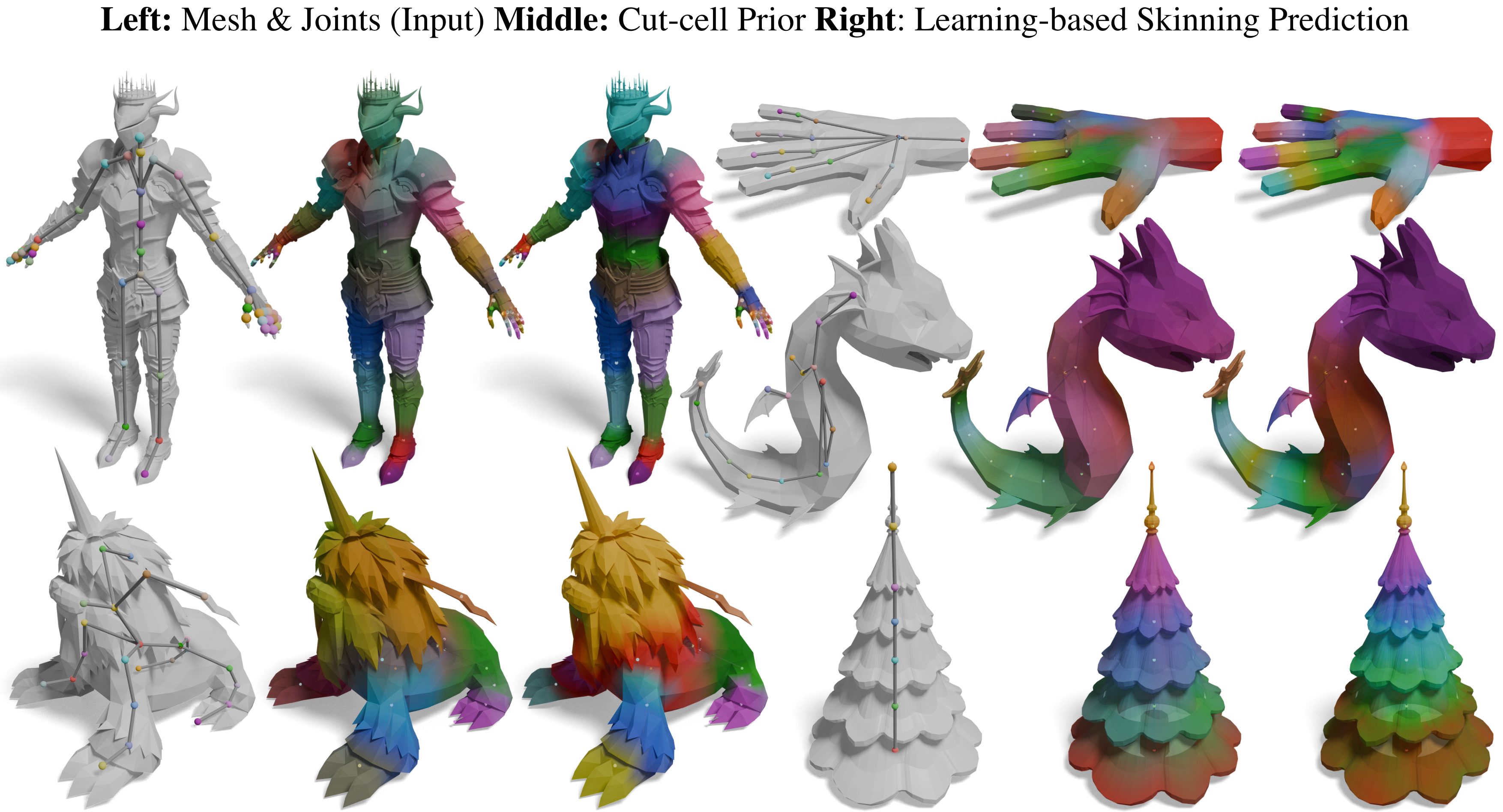}
    \caption{Gallery of paired input meshes and joints, the cut-cell skinning prior, and neural skinning predictions based on the cut-cell prior.}
    \label{fig:cutCellSkinWithoutNetwork}
\end{figure}

\begin{figure}
    \centering
    \includegraphics[width=3.33in]{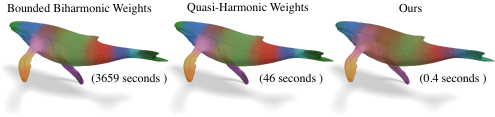}
    \caption{We compare the runtime for skinning prior computation, excluding the time for volumetric meshing and graph construction. Optimization-based methods (e.g., \cite{JacobsonBPS11, WangSolomon21} requires expensive computation, making it challenging to operate on large scale datasets. This motivates our cut-cell skinning prior which only requires to compute the shortest graph geodesics, yielding a more scalable neural skinning pipeline.}
\label{fig:geodesics}
\end{figure}

After constructing the cut-cell graph $\G$, we compute our \emph{skinning prior} based on graph geodesic distances, similar to the method by \citet{DionneL13}.
As described in \reffig{distanceInit}, we sample $N=5$ points for each bone. For each sample point, we identify its closest graph vertex in $\mathcal{G}$. We then define the set of \emph{source vertices} on $\mathcal{G}$ as those vertices that are closest to at least one sample point.
We initialize the distance of each source vertex to its Euclidean distance to the bone, then use Dijkstra's algorithm to compute graph geodesic distances for all remaining non-source vertices in $\G$. This yields the shortest graph geodesic distance $d_{ij}$ from every vertex i to every bone j (\refalg{cutcell_bonedist}). Finally, we transform the shortest graph distance $d_{ij}$ into an unnormalized skinning weight $w_{ij}$ following the formulation in \cite{DionneL13}:
\begin{align}\label{equ:skinning_kernel}
    w_{ij} = \Bigg(\frac{1}{ (1-\alpha) \big(\frac{d_{ij}}{D}\big) + \alpha \big(\frac{d_{ij}}{D}\big)^2} \Bigg)^2 
\end{align}
\begin{wrapfigure}[8]{r}{1.6in}
    \vspace{-6pt}
	\includegraphics[width=\linewidth, trim={10mm 5mm 5mm 5mm}]{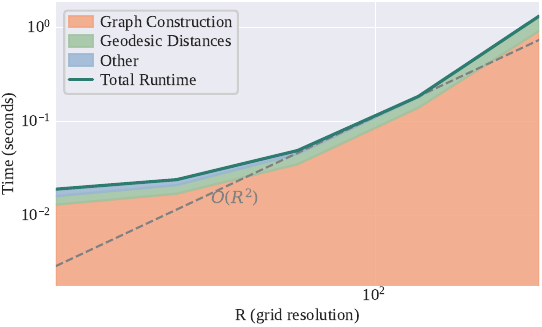}
    \label{fig:running_time}
\end{wrapfigure}
where $\alpha$ is a parameter to control the smoothness of the weight, and $D$ is the bounding box extent to normalize the distance value $d_{ij}$. After computing $w$, the final skinning priors are rescaled at each vertex to ensure partition of unity (see \cite{DionneL13} for more details). In the middle columns of \reffig{cutCellSkinWithoutNetwork}, we show the skinning prior computed from our cut-cell graph geodesics. And the inset shows the running time of our cut-cell prior calculation with respect to the grid resolution. 
Using graph geodesics as the skinning prior yields several orders of magnitude speedups compared to optimization-based methods (see \reffig{geodesics}), making it a viable option to process large datasets on average computation resources.
Compared to the method by \citet{DionneL13} which also uses geodesics as a core subroutine, our method can more robustly separate mesh patches that are nearby under the Euclidean distance, but distant in terms of the geodesic distance (see \reffig{voxelBinding}), leading to a more accurate prior for the neural skinning (see comparisons in \refsec{results}).

\paragraph{Unreachable Vertices}
In rare cases where the mesh contains multiple disconnected components, some mesh vertices in $\mV^\M$ may not be reachable by any source vertices, resulting in constant skinning weights after applying the kernel in \refequ{skinning_kernel}.
To address this, we simply assign each \emph{unreachable} vertex with distance equals to its Euclidean distance to its top-$k$ nearest rig bones before applying \refequ{skinning_kernel}. In practice, we set $k=3$.

\section{Experiments}\label{sec:results}

We incorporate our cut-cell prior into RigNet~\cite{xu2020rignet} and UniRig~\cite{zhang2025unirig} by replacing their original voxel-binding prior~\cite{DionneL13}. We further inject our cut-cell prior into Puppeteer~\cite{song2025puppeteer}, a purely learning-based method that does not rely on geometric skinning priors. All baseline results are reported from their official released checkpoints under the same evaluation protocol. Across all three baselines, the resulting models consistently outperform the original baselines and achieve new state-of-the-art performance.

\subsection{Architectural Modifications}
\textbf{RigNet}~\cite{xu2020rignet} predicts skinning weights using a graph neural network conditioned on a skeleton-aware mesh representation. For each mesh vertex, RigNet computes volumetric geodesic distances to all skeleton bones using voxel-based geodesic binding~\cite{DionneL13}. Bones are ranked by geodesic proximity, and the nearest ones are selected. Their inverse distances, together with the corresponding bone features, are concatenated to form a skeleton-aware per-vertex feature vector, which is processed by a GMEdgeNet to predict the final skinning weights. We integrate our cut-cell skinning prior into RigNet by replacing the voxel-based geodesic distances with graph geodesic distances computed on our cut-cell graph.

\textbf{UniRig}~\cite{zhang2025unirig} predicts skinning weights using a bone--point cross-attention module that models interactions between mesh vertices and skeleton bones. Each bone is encoded by its head and tail positions, while each vertex is encoded using point-wise geometric features extracted by a pretrained point encoder~\cite{wu2024pointtransformer}. The bone--point cross attention scores are concatenated with voxel-based skinning weights computed using geodesic voxel binding~\cite{DionneL13}, and passed through an MLP followed by a softmax to produce the final skinning weights. We integrate our cut-cell skinning prior into UniRig by replacing the voxel-based skinning prior with cut-cell skinning. The skinning network is then trained to predict the residual between the cut-cell skinning and the ground-truth skinning weights.

\textbf{Puppeteer}~\cite{song2025puppeteer} predicts skinning weights using a purely learning-based attention architecture without explicit geometric skinning priors. Mesh vertices are encoded by a point encoder~\cite{yang2024sampart3d} into point-wise features, while bones are encoded into bone features. A series of bone--point cross-attention modules is applied to model vertex--bone interactions. The final skinning weights are obtained by computing cosine similarities between the refined point and bone features, followed by a softmax normalization. To inject our cut-cell skinning prior into Puppeteer, we concatenate the cut-cell skinning weights with the point features after the attention modules. A lightweight MLP fuses our prior and point features before the final cosine similarity computation.

\subsection{Articulation XL-2.0 De-duplication}
While we adopt the Articulation-XL 2.0 dataset commonly used in recent auto-rigging and skinning works, we observe notable redundancy inherited from large-scale web-sourced datasets such as Objaverse~\cite{DeitkeLWNMKFLVG23,DeitkeSSWMVSEKF23}, which can affect quantitative evaluation. In particular, we identify approximately 11K near-duplicate assets, including exact duplicates and meshes that share identical global geometry with only minor variations in vertex or face counts. More critically, we find that 660 of the 1,997 assets in the standard test split also appear in the training set, leading to unintended train–test overlap. To ensure a fair assessment of generalization, we therefore report results on both the original test set and a rigorously deduplicated test split, denoted Articulation-XL (de-duplicated). A detailed description of the dataset statistic and deduplication criteria is provided in the appendix.

\subsection{Evaluation Metrics}
Following RigNet~\cite{xu2020rignet} and Puppeteer~\cite{song2025puppeteer}, we report average $L_1$ distance, precision, and recall. The average $L_1$ distance measures the mean absolute error between predicted and ground-truth skinning weights over all vertices. Precision and recall are computed against weights exceeding a threshold of $0.001$ in the
prediction and ground truth, respectively.

However, we observe that the above-mentioned metrics do not fully reflect skinning quality under animation. Specifically, those metrics do not differentiate whether a vertex is misassigned to a neighboring joint or to a distant joint. In practice, this difference leads to a huge difference in animation quality. In the first row of \reffig{qual_ani_puppeteer}, we show an example where a small skinning weight difference leads to significant visual artifacts because the vertex on the right leg is misassigned to a distant joint, the left leg. 


To capture such ``sticking'' artifacts, a tentative choice is to use a smoothness energy such as the biharmonic
energy~\cite{JoshiMDGS07, JacobsonBPS11, WangSolomon21, DodikSSS24}. However,
as noted by~\cite{BangL18}, artist-authored weights often prioritize semantic part boundaries and material heterogeneity over smoothness. This results in non-smooth weights with sharp transitions, leading to high errors even when measured on the ground truth skinning weight.

\subsubsection{Rest-Post Deformation Error}
Motivated by prior skinning decomposition work~\cite{KavanOMDZ07, LeD12},
we propose a metric that measures skinning quality directly in deformation
space. Due to the absence of artist authored animations, we measure the deformation error as a perturbation from the rest pose. 
For each joint, we rotate it by a fixed angle around a set of canonical axes and apply linear blend skinning to deform
the mesh under the predicted and the ground-truth weights independently.
Precisely, the linear blend skinning function can be written as 
\begin{align}
    \vv_i'(j, \theta, x, \mW) = \sum_{k=1}^m w_{ik} \mT_k(j, \theta, x) \vv_i 
\end{align}
where $w_{ik}$ is the skinning weight between vertex $i$ and joint \textit{k}, \textit{m} is the number of joints, and $\mT_k(j, \theta, x)$ is the transformation matrix at joint $k$ derived from rotating joint $j$ with angle $\theta$ around the $x$ axis using \emph{forward kinematics}. In other words, if $j$ is a parent joint of $k$, rotating joint $j$ will also lead to non-identity $\mT_k$.

Our deformation error measures per-vertex squared Euclidean distance between the vertex locations deformed using the predicted skinning weight $\mW$ and the ground truth skinning weights $\tilde{\mW}^\text{g.t.}$ under a rotation $\theta=30\degree$. For each vertex we report the maximum error over all (joint, axis) pairs, and average across all vertices
\begin{align}
    E_\text{def.} = \frac{1}{n} \sum_{i = 1}^n \argmax_{j, x}\big( \| \vv_i'(j, \theta, x, \mW) - \vv_i'(j, \theta, x, \mW^\text{g.t.}) \|^2 \big)
\end{align}
where $n$ denotes the number of vertices. We normalize each mesh so that its longest bounding-box axis has unit length (1\,m) and report the resulting errors in cm$^2$. We refer to this quantity $E_\text{def}$ as the deformation error in the following. As shown in the first two rows of \reffig{qual_ani_puppeteer}, the deformation error correctly penalizes high-frequency ``sticking'' artifacts that a weight space error, such as the average $L_1$ distance, fails to distinguish: two predictions can have nearly identical $L_1$ values while producing different deformation errors.

\subsection{Quantitative and Qualitative Comparison}
As shown in~\reftab{rignet_quantitative}, \reftab{unirig_quantitative}, and \reftab{puppeteer_quantitative}, injecting our cut-cell prior yields consistent improvements across all baselines, indicating that the prior provides complementary geometric structure information and establishing a new state of the arts. The gains are clearest under the proposed deformation error, where our cut-cell prior yields larger relative improvements than under weight-space metrics such as average $L_1$ error or precision/recall. This is consistent with our earlier observation that the deformation error is more sensitive to the high-frequency ``sticking'' artifacts caused by spurious weights on geodesically distant joints, precisely the failure mode that our geodesic prior is designed to suppress.

\begin{figure}
    \centering
    \includegraphics[width=3.33in]{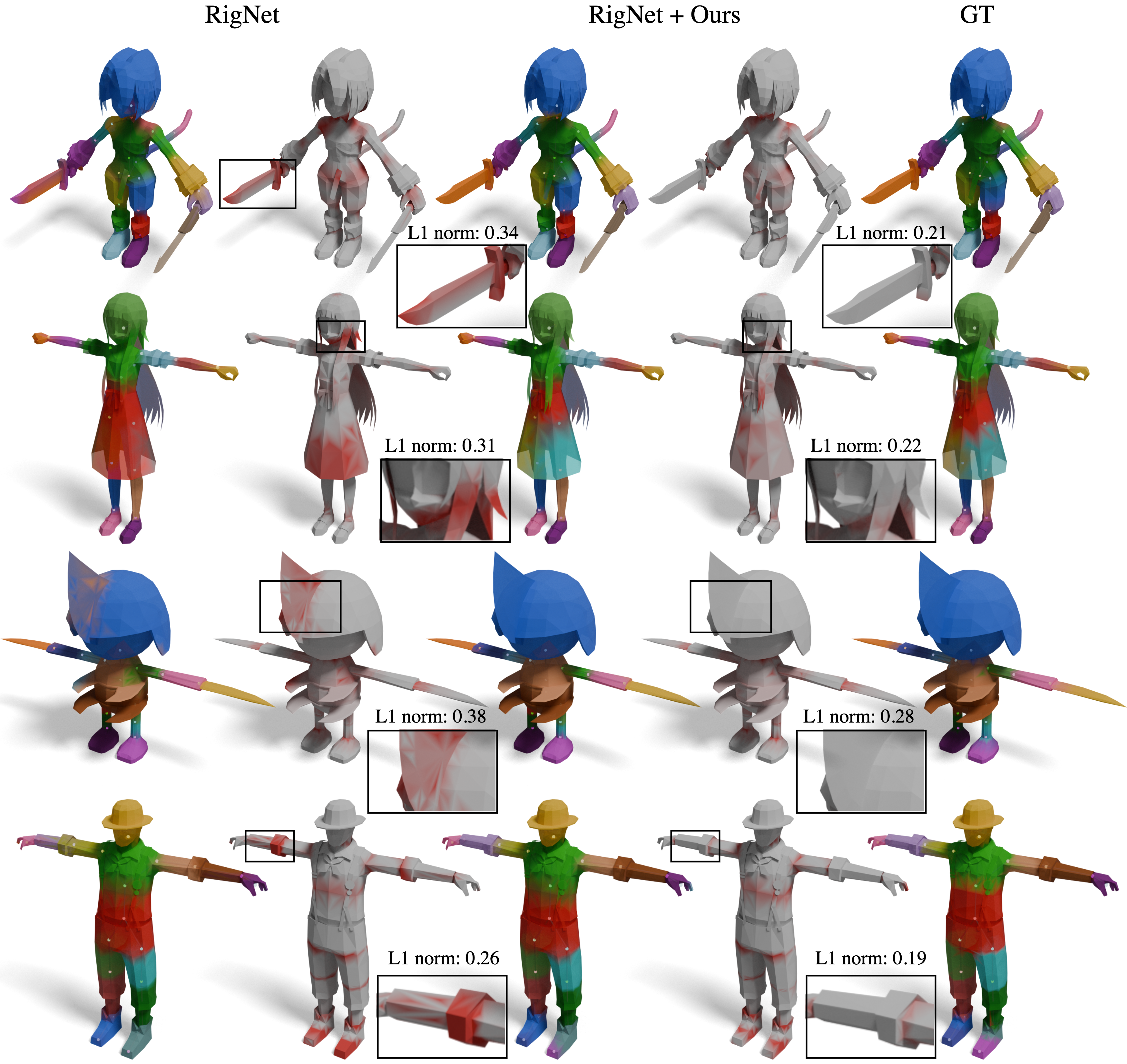}
    \caption{Qualitative comparison with RigNet~\cite{xu2020rignet}. Replacing the geodesic voxel binding with our cut-cell prior better separates geodesically distant but spatially close parts, such as long hair against the shoulder (the second row).}
    \label{fig:qual_rignet}
\end{figure}

\begin{figure}
    \centering
    \includegraphics[width=3.33in]{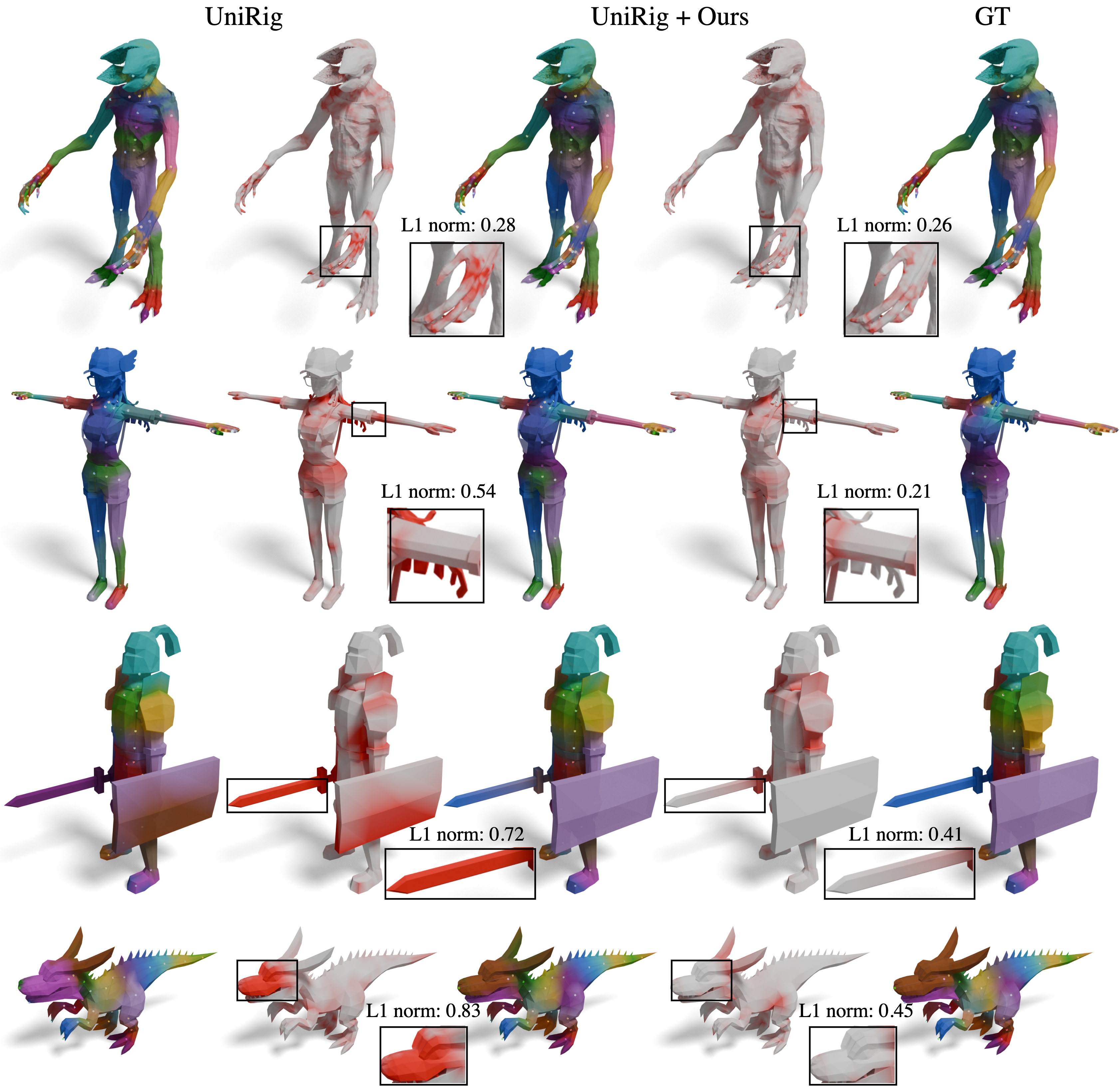}
    \caption{Qualitative comparison with UniRig~\cite{zhang2025unirig}. Compared with using geodesic voxel binding, our cut-cell prior helps the baseline disambiguate fine geometric parts that lie spatially close yet are topologically far apart, such as the creature's long claws (the first row).
    }
    \label{fig:qual_unirig}
\end{figure}

\begin{figure}
    \centering
    \includegraphics[width=3.33in]{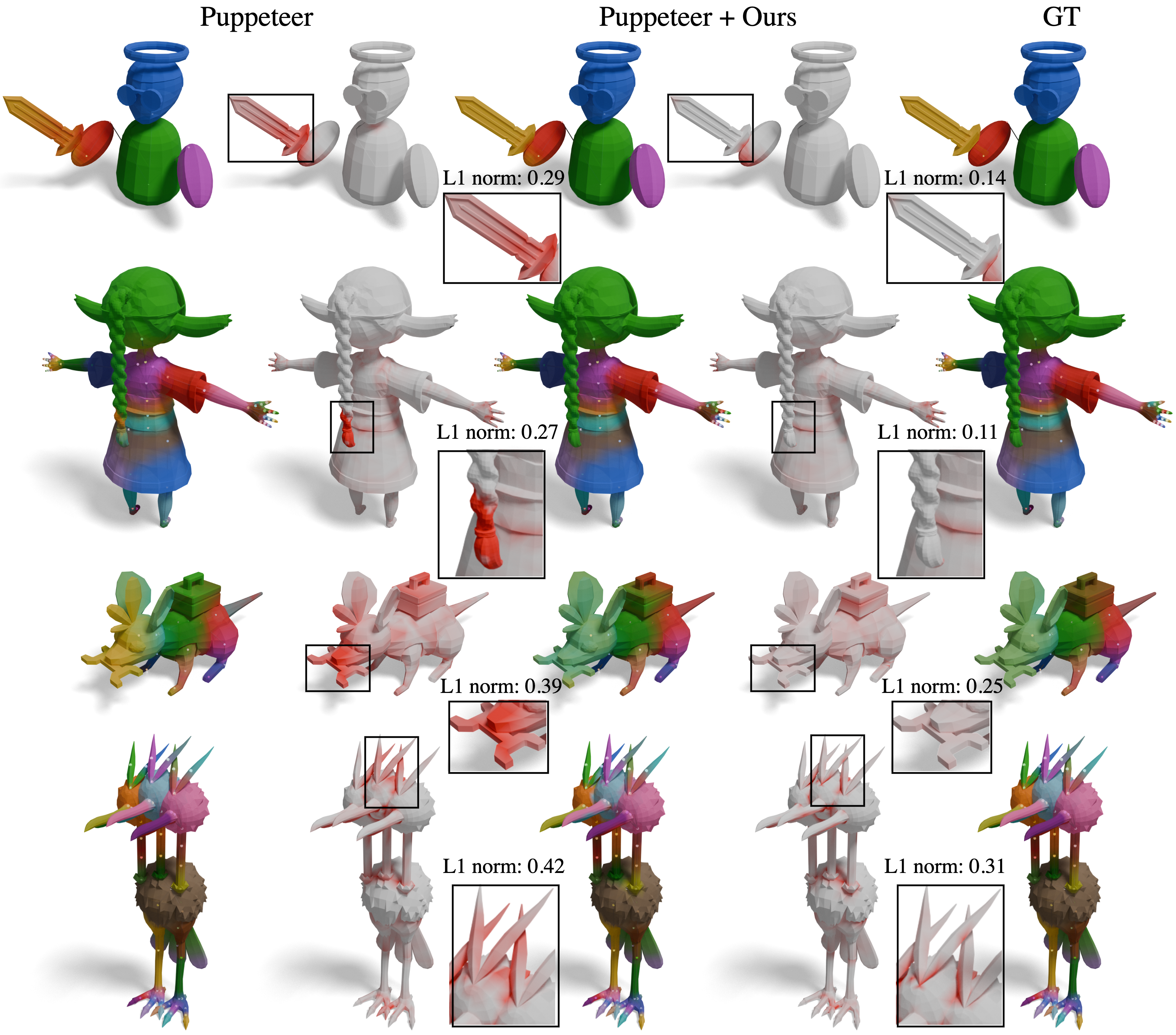}
    \caption{Qualitative comparison with Puppeteer~\cite{song2025puppeteer}. On rare attached objects (e.g., the sword, top row), the baseline fails to correctly capture the semantics of these components, leaking skinning weights from unrelated joints. Our cut-cell prior supplies accurate joint--vertex distance estimates, yielding correctly localized weights.}
    \label{fig:qual_puppeteer}
\end{figure}

\begin{figure}
    \centering
    \includegraphics[width=3.33in]{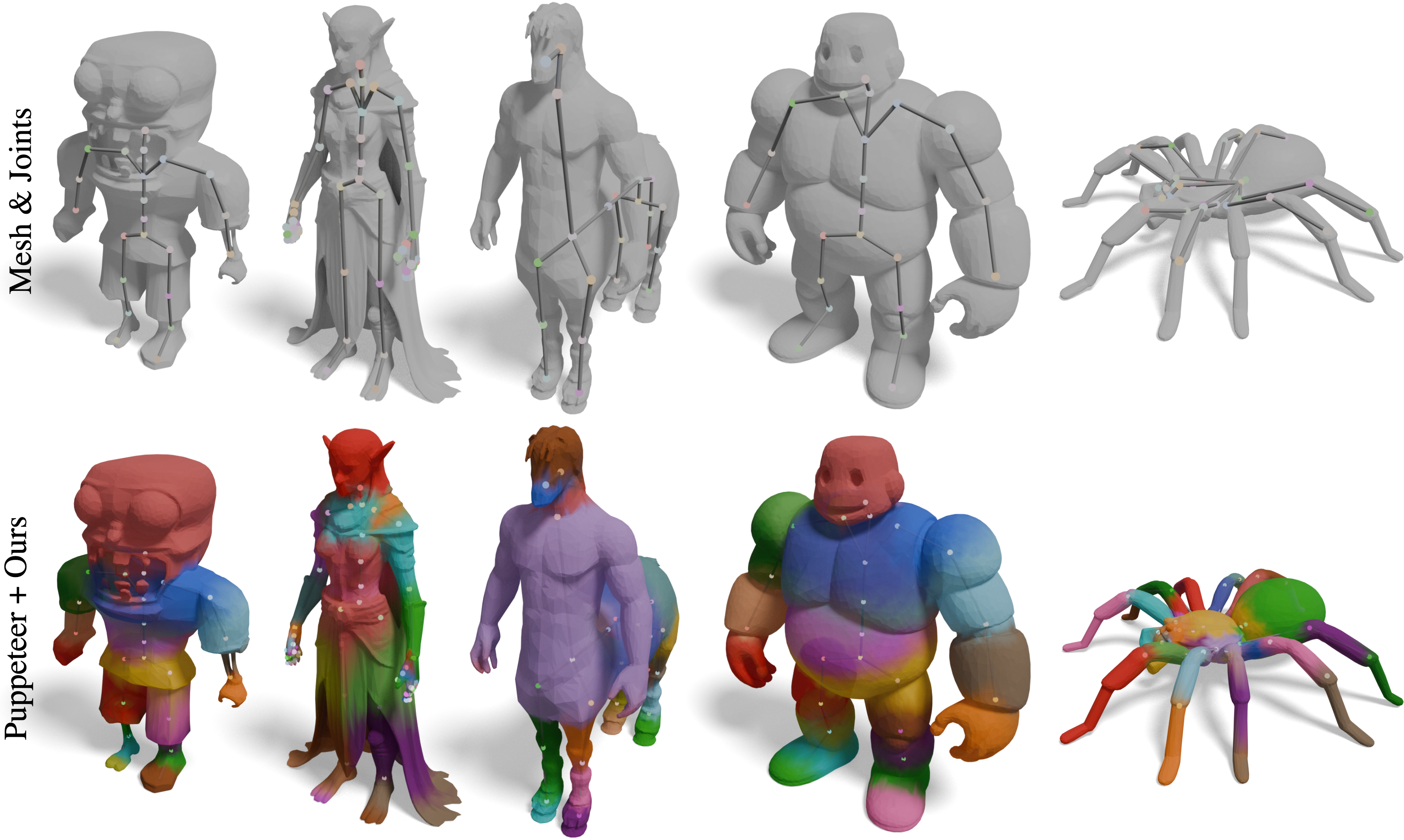}
    \caption{Qualitative results on AI generated characters. Our method generalizes to synthesized meshes beyond the training distribution.}
    \label{fig:qual_ai}
\end{figure}

\begin{figure}
    \centering
    \includegraphics[width=3.33in]{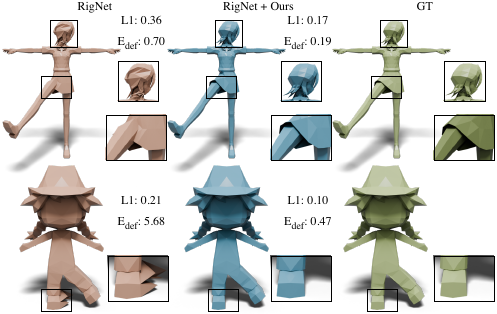}
    \vspace{-2mm}
    \caption{Qualitative comparison of animation results with RigNet~\cite{xu2020rignet}. The baseline produces visible ``sticking''
artifacts at the hair (the first row) and feet (the second row). Our cut-cell prior corrects these errors.
    }
    \label{fig:qual_ani_rignet}
\end{figure}

\begin{figure}
    \centering
    \includegraphics[width=3.33in]{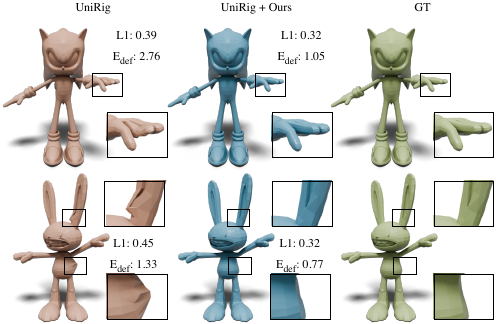}
    \vspace{-2mm}
    \caption{Qualitative comparison of animation results with UniRig~\cite{zhang2025unirig}.The baseline produces unnatural deformations at slender extremities: the middle finger is dragged along with the index finger (top row), and the rabbit's ear root and hip exhibit jarring kinks. Our cut-cell prior recovers natural, GT-consistent
deformations in both cases.} 
    \label{fig:qual_ani_unirig}
\end{figure}

\begin{figure}
    \centering
    \includegraphics[width=3.33in]{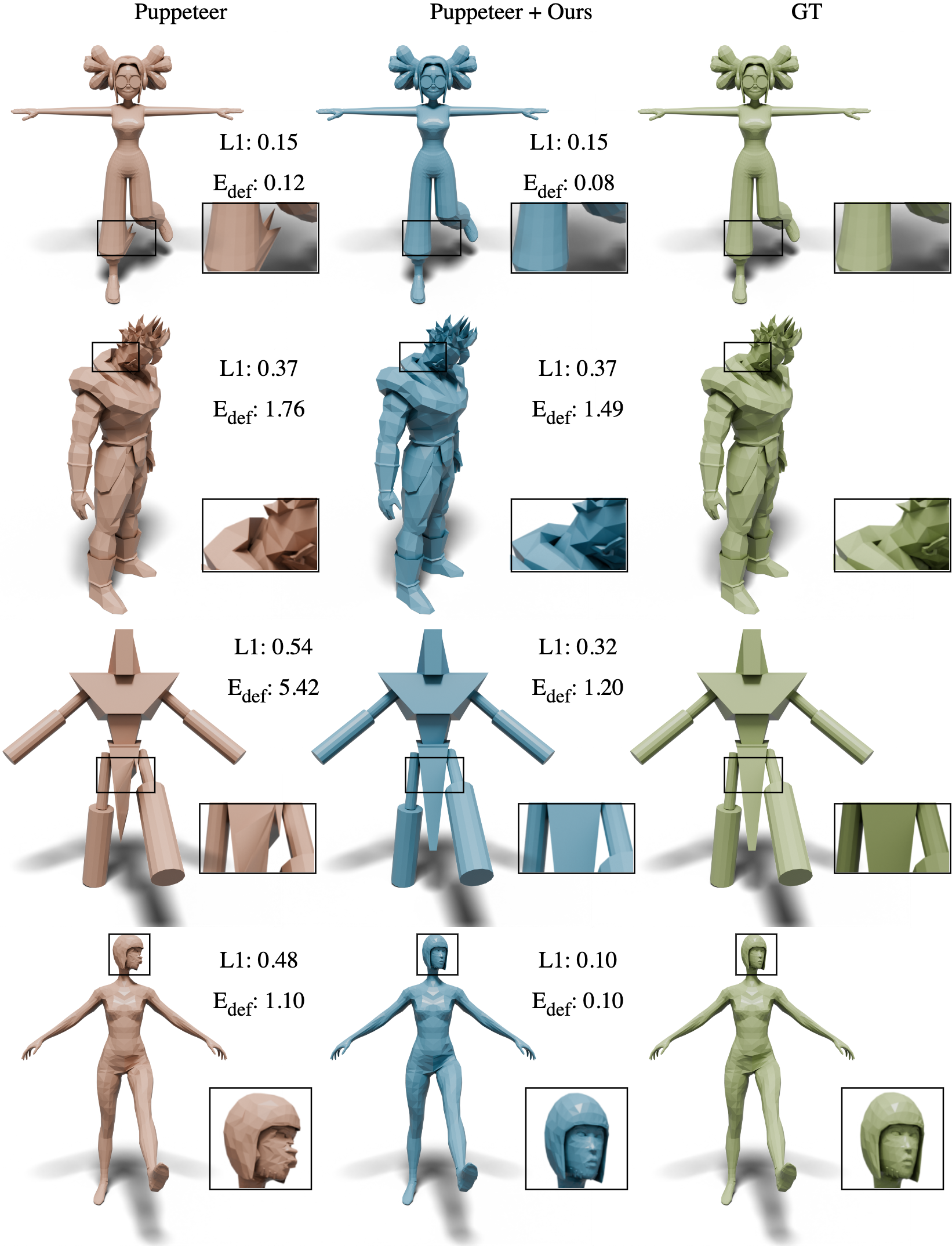}
    \vspace{-5mm}
    \caption{Qualitative comparison of animation results against Puppeteer~\cite{song2025puppeteer}. In the first two rows, the two methods produce nearly identical $L_1$ values, yet our cut-cell prior yields visibly cleaner deformations without "sticking" artifacts and a noticeably lower $E_{\text{def}}$. In the bottom two rows, our prior improves both metrics, with the deformation error reduced by a larger margin than $L_1$.}
    \vspace{-2mm}
    \label{fig:qual_ani_puppeteer}
\end{figure}

We provide qualitative comparisons of predicted skinning weights in Figs.~\ref{fig:qual_rignet},~\ref{fig:qual_unirig}, and~\ref{fig:qual_puppeteer} for RigNet, UniRig, and Puppeteer, respectively. Across all methods, models augmented with our cut-cell prior produce skinning weights that better respect geometric locality and semantic part boundaries. We also present skinning prediction results in \reffig{qual_ai} on meshes generated by a text-to-3D model, demonstrating that our method generalizes to synthesized meshes beyond the training distribution. We further evaluate animation quality in Figs.~\ref{fig:qual_ani_rignet},~\ref{fig:qual_ani_unirig}, and~\ref{fig:qual_ani_puppeteer}, where improved skinning accuracy translates into visibly more stable deformations under large joint rotations and reduced artifacts in spatially adjacent regions.\textit{Please also refer to the supplementary video and HTML for dynamic animation results.}

\begin{table}[htb!]
    \centering
    \caption{We swap out the geodesic voxel binding \cite{DionneL13} component from the RigNet~\cite{xu2020rignet} and show improvements on the ModelsResource dataset they trained on.}
    \vspace{-2mm}
    \resizebox{0.8\linewidth}{!}{%
    \begin{tabular}{c c c c c}
    \toprule
         & avg L1 $\downarrow$ & Precision $\uparrow$ & Recall $\uparrow$ & $E_{def} \downarrow$  \\
     \midrule
        RigNet & 0.432 & 0.803 & 0.794 & 8.58 \\
        RigNet + Ours & \textbf{0.367} & \textbf{0.850} & \textbf{0.798} & \textbf{6.06} \\
        \midrule
        Improvement & 15.05\% & 5.85\% & 0.50\% & 29.37\% \\
    \bottomrule
    \end{tabular}
    }
    \label{tab:rignet_quantitative}
    \vspace{-4mm}
\end{table}

\begin{table}[htb!]
\centering
\caption{We replace the geodesic voxel binding~\cite{DionneL13} in UniRig~\cite{zhang2025unirig} with our cut-cell skinning prior. Results are reported on the Articulation-XL2.0 dataset and its de-duplicated variant.}
\vspace{-2mm}
\label{tab:unirig_quantitative}
\resizebox{\linewidth}{!}{
\begin{tabular}{l cccc cccc}
\toprule
& \multicolumn{4}{c}{Articulation-XL2.0} 
& \multicolumn{4}{c}{Articulation-XL2.0 (de-duplicated)} \\
\cmidrule(lr){2-5} \cmidrule(lr){6-9}
Method 
& avg L1 $\downarrow$ & Prec. $\uparrow$ & Recall $\uparrow$ & $E_{def}$ $\downarrow$
& avg L1 $\downarrow$ & Prec. $\uparrow$ & Recall $\uparrow$ & $E_{def} \downarrow$\\
\midrule
UniRig
& 0.747 & 0.746 & 0.673 & 21.25 
& 0.768 & 0.736 & 0.663 & 25.29 \\
UniRig + Ours 
& \textbf{0.386} & \textbf{0.820} & \textbf{0.850} & \textbf{10.91}
& \textbf{0.447} & \textbf{0.794} & \textbf{0.829} & \textbf{14.18} \\
\midrule
Improvement
& 48.33\% & 9.92\% & 26.30\% & 48.66\%
& 41.80\% & 7.88\% & 25.04\% & 43.93\% \\
\bottomrule
\end{tabular}
}
\end{table}

\begin{table}[hbtp]
\centering
\caption{Effect of injecting our cut-cell skinning prior into Puppeteer~\cite{song2025puppeteer}. Results are reported on the Articulation-XL2.0 dataset and its de-duplicated variant.}
\vspace{-3mm}
\label{tab:puppeteer_quantitative}
\resizebox{\columnwidth}{!}{%
\begin{tabular}{l cccc cccc}
\toprule
& \multicolumn{4}{c}{Articulation-XL2.0} 
& \multicolumn{4}{c}{Articulation-XL2.0 (de-duplicated)} \\
\cmidrule(lr){2-5} \cmidrule(lr){6-9}
Method 
& avg L1 $\downarrow$ & Prec. $\uparrow$ & Recall $\uparrow$ & $E_{def} \downarrow$
& avg L1 $\downarrow$ & Prec. $\uparrow$ & Recall $\uparrow$ & $E_{def} \downarrow$ \\
\midrule
Puppeteer
& 0.335 & 0.876 & \textbf{0.740} & 5.949 
& 0.377 & 0.859 & \textbf{0.735} & 7.763 \\
Puppeteer + Ours
& \textbf{0.320} & \textbf{0.896} & 0.724 & \textbf{5.333} 
& \textbf{0.368} & \textbf{0.877} & 0.712 & \textbf{7.453} \\
\midrule
Improvement
& +4.48\% & +2.28\% & -2.16\% & +10.36\%
& +2.39\% & +2.10\% & -3.13\% & +3.99\% \\
\bottomrule
\end{tabular}}
\end{table}

\subsection{Runtime and Accuracy of the Cut-cell Prior}
\begin{figure}
    \centering
    \includegraphics[width=3.33in]{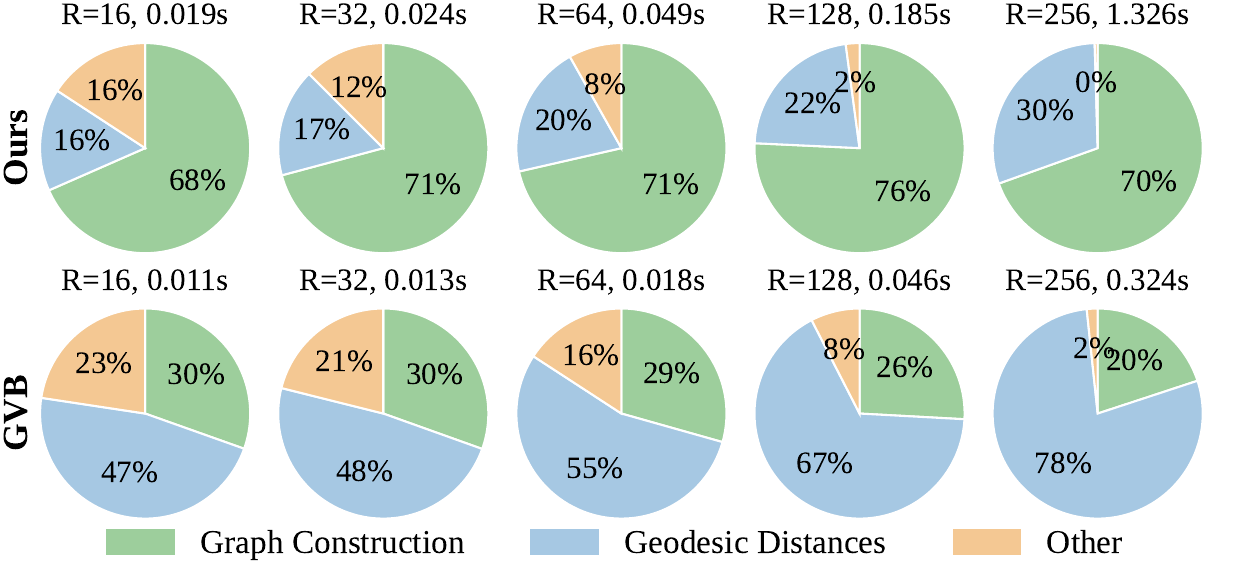}
    \caption{Runtime breakdown of the cut-cell prior and geodesic voxel binding computation, where R denotes the grid resolution.}
    \label{fig:runtime}
\end{figure}

\begin{figure}
    \centering
    \includegraphics[width=3.33in]{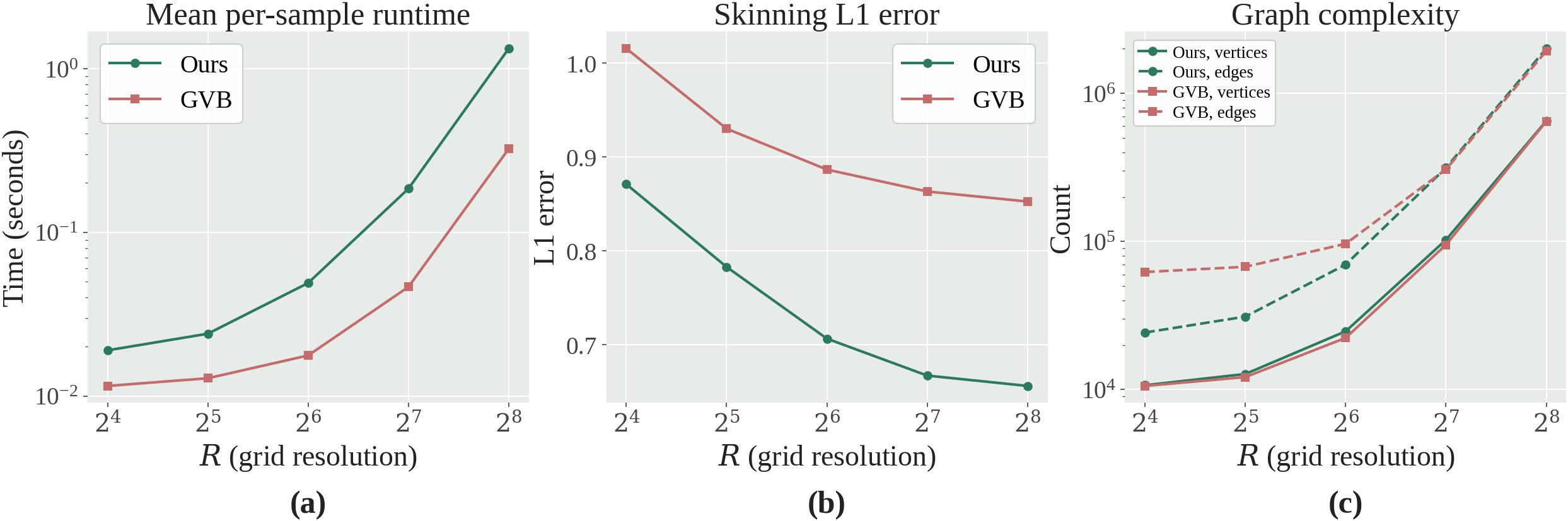}
    \vspace{-3mm}
    \caption{Runtime, $l1$ error, and graph complexity comparison with Geodesic Voxel Binding~\cite{DionneL13}. Although our cutcell prior  incurs a small computational overhead per resolution, it reaches higher accuracy than Geodesic Voxel Binding at a much coarser grid (e.g., $R=2^6$ vs.\ $R=2^8$), yielding both shorter overall runtime and more precise skinning priors.}
    \label{fig:gvb}
\end{figure}

We compare our cut-cell prior against the widely used Geodesic Voxel
Binding~\cite{DionneL13} on the full Articulation-XL 2.0 test set. All
experiments are conducted on a Linux workstation with an Intel i9-14900K
CPU and 24~GB of RAM. \reffig{gvb} reports runtime, accuracy, and graph complexity across grid resolutions and \reffig{runtime} shows runtime breakdown.

Our method achieves substantially higher accuracy at every resolution. Although it incurs a small computational overhead per resolution, it reaches higher accuracy than Geodesic Voxel Binding at a much coarser grid (e.g., $R=64$ vs.\ $R=256$), yielding both shorter overall runtime and more precise skinning priors.

We further benchmark our cut-cell graph construction against fTetWild~\cite{HuSWZP20} and Mandoline~\cite{TaoBFL19} on 10 samples from the Articulation-XL 2.0 test set~\cite{song2025magic}, limited by the baselines' high memory overhead on large meshes. Our method outperforms them by orders of magnitude across all resolutions (\reftab{graph_construction}).

\begin{table}[htbp]
\centering
\caption{Graph construction time comparison (in seconds). Values
in parentheses indicate the speedup factor of our method over each
baseline.}
\vspace{-4mm}
\resizebox{0.8\linewidth}{!}{
\begin{tabular}{lccccc}
\toprule
Grid Resolution & 8 & 16 & 32 & 64 & 128 \\
\midrule
fTetWild~\cite{HuSWZP20} & 5.72\,(636$\times$) & 6.43\,(643$\times$) & 6.84\,(570$\times$) & 13.37\,(461$\times$) & 75.58\,(548$\times$) \\
Mandoline~\cite{TaoBFL19} & 0.12\,(13$\times$) & 0.19\,(19$\times$) & 0.45\,(38$\times$) & 2.50\,(86$\times$) & 24.69\,(179$\times$) \\
Ours & \textbf{0.009} & \textbf{0.010} & \textbf{0.012} & \textbf{0.029} & \textbf{0.138} \\
\bottomrule
\end{tabular}
}
\label{tab:graph_construction}
\vspace{-4mm}
\end{table}
\section{Discussion \& Conclusion}
We introduced the cut-cell prior: a fast, robust, graph-based approximation of volumetric geodesic distances on in-the-wild meshes. Integrated into state-of-the-art neural skinning models, it consistently improves generalization, showing that geometric reasoning and semantic learning are complementary for neural skinning.

\begin{figure}
    \centering
    \includegraphics[width=3.33in]{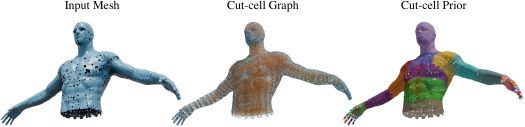}
    \vspace{-4mm}
    \caption{Robustness on polygon-soup inputs. The input mesh (left) is open and non-watertight, with holes scattered across the surface. Our cut-cell graph (middle) is built robustly and the resulting skinning prior (right) yields part-consistent regions.}
    \label{fig:robust_analyze}
\end{figure}

Our method builds the cut-cell graph based on fast winding number queries and axis-aligned ray casts, which are robust to non-watertight meshes, noisy
geometry, and polygon soups (\reffig{robust_analyze}). However, the
winding-number-based inside/outside classification assumes that the input
mesh roughly approximates a closed volume, and remains sensitive to
inverted triangles and thin shells that do not bound a solid region.
Encouraging consistent facet orientation through techniques such as~\cite{XuDWXCJGWT23, solidShell} could further improve robustness in
these cases.

\begin{figure}
    \centering
    \includegraphics[width=3.33in]{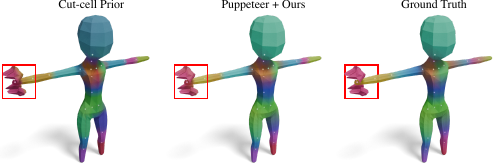}
    \vspace{-4mm}
    \caption{Failure case on l2 fallback. The rock ring (red box) consists of components disconnected from the body mesh, so its vertices cannot reach any joint along the cut-cell graph. Our method falls back to Euclidean distance, assigning each vertex to its three nearest joints in space. The downstream network corrects this through its semantic understanding of the shape.}
    \label{fig:failure_case}
\end{figure}

For meshes that contain disconnected components unreachable from any joint along the cut-cell graph, we fall back to Euclidean distance, assigning each
unreachable vertex to its nearest joints in space. This fallback can
incorrectly bind vertices to unrelated joints that happen to be
spatially close, as shown in \reffig{failure_case}. The downstream network corrects this through its semantic understanding of the shape. Our cut-cell graph has more vertices than cage-based representations, since cages merge spatially close but distinct parts, this finer structure yields a stronger geometric prior at the cost of moderately longer geodesic computation. While increasing the grid resolution empirically improves accuracy by enriching cut-cell graph connectivity, future exploration on using quadrature or adding extra edge connectivity (similar to \cite{WangFWXH17}) could ensure the geodesic approximation converges to true volumetric geodesics under refinement.

\clearpage
\bibliographystyle{ACM-Reference-Format}
\bibliography{sections/reference}

\clearpage

\clearpage
 \appendix
\section{Data Audit}



This section details the comprehensive data analysis performed to validate the integrity of the Articulation-XL 2.0 dataset prior to our skinning experiments. As noted in Section 4.2, Articulation-XL 2.0 was selected for its scale (48K assets) and categorical diversity (humanoid, anthropomorphic, animal, etc.). However, to ensure the validity of our proposed cutcell-based skinning prior, we identified the need to rigorously assess the dataset for internal redundancies and train-test overlaps.

\subsection{Motivation for Data Audit}

To ensure the integrity of our evaluation, we introduce a dataset audit process comprising two distinct protocols: (i) \textbf{Data Leakage Analysis} and (ii) \textbf{Duplicate Data Detection}. Data leakage—the presence of training samples within the test set—can result in overly optimistic metrics that reflect overfitting rather than true generalization. Furthermore, duplicate data causes training inefficiency and biases the model toward frequent shapes. 
Crucially, our analysis extends beyond identical copies to encompass \textit{near-duplicates}: assets sharing nearly identical global shape despite minor geometric and topological variations.
To ensure robust benchmarking, we evaluate on both the complete test set and a curated subset that excludes these overlapping assets. This comprehensive audit enhances training efficiency, mitigates bias, and prevents misevaluation in future rigging frameworks.

Our analysis yielded significant findings. We identified \textbf{660 out of 1,997} ($\sim$33.0\%) test set assets as near-duplicates of training data. Furthermore, across the entire Articulation-XL 2.0 dataset, we found approximately \textbf{11K out of 48K} ($\sim$23\%) assets to be near-duplicates. The detailed audit methodology is described in the following subsection.

\subsection{Data Audit Protocols}

To efficiently process the large-scale dataset $\mathcal{S} = \{s_1, \dots, s_N\}$, we employ a hierarchical two-stage approach combining coarse skeletal filtering with fine-grained geometric verification. 

\textbf{Geometric Definitions.} 
To enable comparison independent of scale, all assets are normalized to a canonical unit cube. For robust handling of topological variations, we sample points on the mesh faces to generate a dense point cloud $\mathcal{P}_i$ for every asset $s_i$, rather than relying on raw vertices. We quantify the similarity between any two point sets $\mathcal{A}$ and $\mathcal{B}$ (representing either joints or surface points) using the symmetric Chamfer Distance ($D_{\text{CD}}$):
\begin{equation}
    D_{\text{CD}}(\mathcal{A}, \mathcal{B}) = \frac{1}{|\mathcal{A}|} \sum_{a \in \mathcal{A}} \min_{b \in \mathcal{B}} \|a - b\|^2_2 + \frac{1}{|\mathcal{B}|} \sum_{b \in \mathcal{B}} \min_{a \in \mathcal{A}} \|b - a\|^2_2
\end{equation}

\textbf{Stage 1: Coarse Skeletal Filtering.} 
We first compute the distance between the joint locations ($\mathcal{J}$) of asset pairs, denoted as $D^J_{ij} = D_{\text{CD}}(\mathcal{J}_i, \mathcal{J}_j)$. We define the set of \textit{skeletal candidates}, $\mathcal{C}_{\text{skel}}$, by retaining pairs with skeletal similarity below a threshold $T^J = 0.01$:
\begin{equation}
    \mathcal{C}_{\text{skel}} = \left\{ (s_i, s_j) \mid s_i, s_j \in \mathcal{S}, i \neq j, D^J_{ij} < T^J \right\}
\end{equation}
This step rapidly filters out distinct poses, significantly reducing the search space.

\begin{figure}
     \centering
     \includegraphics[width=0.99\linewidth]{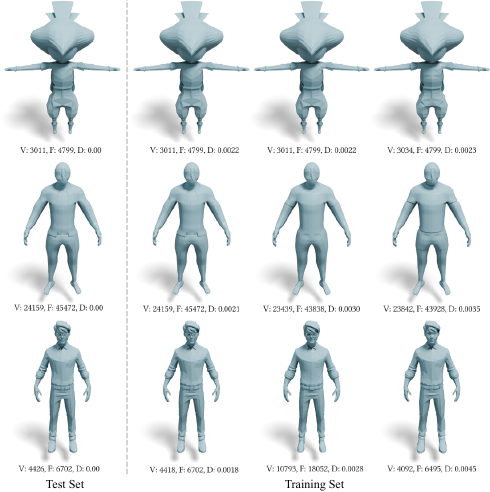}
     \caption{Examples of data leakage in Articulation-XL 2.0.
     V, F and D represent number of vertices, number of faces and the chamfer distance from the test set asset respectively. 
     }
     \label{fig:axl_data_leakage}
\end{figure}

\textbf{Stage 2: Geometric Verification.} 
For pairs in $\mathcal{C}_{\text{skel}}$, we compute the dense vertex-to-vertex Chamfer Distance using the sampled point clouds, $D^V_{ij} = D_{\text{CD}}(\mathcal{P}_i, \mathcal{P}_j)$. A pair is classified as a \textit{near-duplicate} if it belongs to the set $\mathcal{C}_{\text{dup}}$:
\begin{equation}
    \mathcal{C}_{\text{dup}} = \left\{ (s_i, s_j) \in \mathcal{C}_{\text{skel}} \mid D^V_{ij} < T^V \right\}
\end{equation}
where $T^V = 0.01$.
To get a sense of typical chamfer distance magnitudes in thr near-duplicate set, please see \reffig{axl_data_leakage}

\noindent \textbf{Audit Workflows.}
We utilize this hierarchical framework in two specific contexts. 
First, for cross-dataset leakage analysis (Algorithm~\ref{alg:data_leakage_analysis}), we compare a source set $\mathcal{S}_A$ (e.g., Test) against a target set $\mathcal{S}_B$ (e.g., Train) to identify matches $\mathcal{C}_{\text{dup}}$. 
Second, for near-duplicate data detection (Algorithm~\ref{alg:dataset_deduplication}), we process the complete dataset to find redundant groups. 
We define a pairwise similarity score $S_{ij} = \exp(-D^V_{ij}/T^V)$ for candidates and apply DBSCAN clustering on the dissimilarity $1 - S_{ij}$ (with DBSCAN distance threshold $\epsilon=0.5$). 
For each resulting cluster, a single representative is retained. 
Our proposed protocols allow control over audit strictness by exposing $T^J$ and $T^V$ as tunable  parameters.

\reffig{axl_data_leakage} illustrates representative instances of data leakage and redundancy within the Articulation-XL 2.0 dataset. The top row depicts a test asset alongside its training counterparts, which exhibit either identical geometry or negligible topological variations. Subsequent rows present test-train pairs arranged by ascending Chamfer Distance, visually demonstrating a spectrum of near-duplicates, ranging from minor topological deviations to noticeable shape variances. Notably, the example in the third column of the final row highlights a critical edge case: an asset with significantly different mesh topology that retains high geometric similarity. This observation is particularly relevant for point-cloud-based methods like UniRig and Puppeteer, which are agnostic to mesh connectivity and would effectively perceive these assets as almost identical.


\begin{algorithm}[tbp]
\caption{Dataset Leakage Analysis}
\label{alg:data_leakage_analysis}
\DontPrintSemicolon
\KwIn{Source Set $\mathcal{S}_A$, Target Set $\mathcal{S}_B$}
\KwOut{Matching pairs $\mathcal{C}_{\text{dup}}$}

\tcp{Transform and sample point clouds}
$\mathcal{P}^A_i \leftarrow \textsc{TransformAndSample}(\text{mesh}_i)$ for all $s_i \in \mathcal{S}_A$\;
$\mathcal{P}^B_j \leftarrow \textsc{TransformAndSample}(\text{mesh}_j)$ for all $s_j \in \mathcal{S}_B$\;

Initialize $\mathcal{C}_{\text{dup}} \leftarrow \emptyset$, $\mathcal{C}_{\text{skel}} \leftarrow \emptyset$\;

\tcp{Stage 1: Coarse Joint-based Filtering }
\For{$s_i \in \mathcal{S}_A, s_j \in \mathcal{S}_B$}{
  $D^J_{ij} \leftarrow D_{\text{CD}}(\mathcal{J}^A_i, \mathcal{J}^B_j)$\;
  \If{$D^J_{ij} < T^J$}{
    Add $(i, j)$ to $\mathcal{C}_{\text{skel}}$\;
  }
}
\tcp{Stage 2: Fine Geometric Verification }
\For{$(i, j) \in \mathcal{C}_{\text{skel}}$}{
  $D^V_{ij} \leftarrow D_{\text{CD}}(\mathcal{P}^A_i, \mathcal{P}^B_j)$\;
  \If{$D^V_{ij} < T^V$}{
     Add $(s_i, s_j)$ to $\mathcal{C}_{\text{dup}}$\;
  }
}

\Return{$\mathcal{C}_{\text{dup}}$}
\end{algorithm}

\begin{algorithm}[tbp]
\caption{Duplicate Data Detection}
\label{alg:dataset_deduplication}
\DontPrintSemicolon
\KwIn{Dataset $\mathcal{S} = \{s_1, \dots, s_N\}$}
\KwOut{Unique assets $\mathcal{U}$, Duplicate groups $\mathcal{G}$}

\tcp{Transform and sample point clouds}
$\mathcal{P}_i \leftarrow \textsc{TransformAndSample}(\text{mesh}_i)$ for all $s_i \in \mathcal{S}$\;
Initialize $\mathcal{C}_{\text{skel}} \leftarrow \emptyset$, Distance Matrix $\mathbf{D}$\;

\tcp{Stage 1: Coarse Joint-based Filtering}
\For{$1 \le i < j \le N$}{
  $D^J_{ij} \leftarrow D_{\text{CD}}(\mathcal{J}_i, \mathcal{J}_j)$\;
  \If{$D^J_{ij} < T^J$}{
    Add $(i, j)$ to $\mathcal{C}_{\text{skel}}$\;
  }
}

\tcp{Stage 2: Fine Geometric Verification}
\For{$(i, j) \in \mathcal{C}_{\text{skel}}$}{
  $D^V_{ij} \leftarrow D_{\text{CD}}(\mathcal{P}_i, \mathcal{P}_j)$\;
  Fill sparse matrix $\mathbf{D}_{ij} \leftarrow D^V_{ij}$\;
}

\tcp{Convert dist to similarity}
$\mathbf{S} \leftarrow \exp(-\mathbf{D} / T^V)$


\tcp{Clustering}
$\text{labels} \leftarrow \textsc{DBSCAN}(\text{metric} = 1 - \mathbf{S}, \epsilon=0.5)$\;

$\mathcal{U} \leftarrow \{s_i \mid \text{labels}[i] = -1\}$ \tcp{Unclustered samples}
$\mathcal{G} \leftarrow \emptyset$\;
\For{cluster $C$ in \textsc{GetGroups}(labels)}{
   $\mathcal{U} \leftarrow \mathcal{U} \cup \{C[0]\}$ \tcp*{Representative}
   $\mathcal{G} \leftarrow \mathcal{G} \cup C[1{:}]$ \tcp*{Duplicates}
}
\Return{$\mathcal{U}, \mathcal{G}$}
\end{algorithm}

\end{document}